\documentclass[a4paper,12pt]{article}
\usepackage{graphicx,color, pifont,psfrag}
\usepackage{amsmath,amssymb,a4wide,epsfig,mathtools}
\date{\today}
 \normalsize

\newcommand{\bmat}{\left(\begin{array}}
\newcommand{\emat}{\end{array}\right)}
\newcommand{\be}{\begin{equation}}
\newcommand{\ee}{\end{equation}}
\newcommand{\bea}{\begin{eqnarray}}
\newcommand{\eea}{\end{eqnarray}}

\def\threebody{B\to K_1\gamma \to (K \pi\pi) \gamma}
\def\Bbar{\overline{B}}
\def\Kbar{\overline{K}}
\def\lamg{\lambda_\gamma}
\def\jay{\mathcal{J}}
\def\see{\mathcal{C}}
\def\lsim{\raise0.3ex\hbox{$\;<$\kern-0.75em\raise-1.1ex\hbox{$\sim\;$}}}
\def\gsim{\raise0.3ex\hbox{$\;>$\kern-0.75em\raise-1.1ex\hbox{$\sim\;$}}}

\def\no{\nonumber\\}

\definecolor{grey}{rgb}{0.3,0.3,0.3}

\begin{document}
\bibliographystyle{unsrt}
\renewcommand{\thefootnote}{\fnsymbol{footnote}}
\rightline{LPT-ORSAY-10-91} \rightline{LAL-10-209}
\vspace{.3cm} 
{\Large
\begin{center}
{\bf Determining the photon polarization of the $b\to s\gamma$ using the $B\to  K_1(1270) \gamma\to (K\pi\pi)\gamma$ decay}
\end{center}}
\vspace{.3cm}

\begin{center}
E. Kou$^a$, A. Le Yaouanc$^b$, A. Tayduganov$^{a,b}$\\
\vspace{.3cm}\small
\emph{$^a$ 
Laboratoire de l'Acc\'el\'erateur Lin\'eaire,
Universit\'e Paris-Sud 11, CNRS/IN2P3 (UMR 8607)} \\
\emph{91405 Orsay, France}

\emph{$^b$ Laboratoire de Physique Th\'eorique, 
CNRS/Univ. Paris-Sud 11 (UMR 8627)}\\
\emph{91405 Orsay, France}
\end{center}

\vspace{.3cm}
\hrule \vskip 0.3cm
\begin{center}
\small{\bf Abstract}\\[3mm]
\end{center}
Recently the radiative $B$ decay to the strange axial-vector mesons, $B\to K_1(1270) \gamma$, has been observed with rather large branching ratio. This process is particularly interesting as the subsequent $K_1$ decay into its three body final state allows us to determine the polarization of the photon, which is mostly left- (right-)handed for $\overline{B} (B)$ in the SM while various new physics models predict additional right- (left-)handed components. A new method is proposed to determine the polarization, exploiting the full Dalitz plot distribution, which seems to reduce significantly the statistical errors.
This polarization measurement requires however a detailed knowledge of the $K_1 \to K \pi \pi$ strong interaction decays, namely, the various partial wave amplitudes into the several possible quasi two-body channels, as well as their relative phases. 
The pattern of partial waves is especially complex for the $K_1(1270)$. We attempt to obtain the information through the combination of an experimental input and a theoretical one, provided by the $^3P_0$ quark-pair-creation model.
\begin{minipage}[h]{14.0cm}
\end{minipage}
\vskip 0.3cm \hrule \vskip 1.2cm
%
\section{Introduction}\label{sec:1}

The $b\to s\gamma$ process has been playing important roles to understand the electro-weak interaction of the Standard Model (SM). The Glashow-Iliopoulos-Maiani  mechanism shows that in the SM, the flavour changing neutral current such as $b\to s\gamma$ is forbidden at the tree level but only occurs through a loop level diagram. Inside of the loop, heavy particles, much heavier than $b$ quark, can propagate. Therefore, the $b\to s\gamma$ process can be used to probe indirectly such heavy particles, namely top quarks in the case of SM or yet unknown particles introduced by given models beyond the SM. 

By now, the branching ratio of the inclusive $B\to X_s\gamma$ process is measured at quite a high precision ($Br(B\to X_s\gamma)_{\text{exp}}=(3.55\pm0.24\pm0.09)\times10^{-4}$~\cite{HFAG}. The SM theoretical predictions for this observable is obtained at the next-to-next-to-leading order in QCD ($Br(B\to X_s\gamma)_{\text{th}}=(3.15\pm0.23)\times10^{-4}$~\cite{Misiak})  and they are relatively in good agreement with the experimental value.  However, these  predictions have theoretical uncertainties coming from the CKM matrix element as well as various kinds of QCD corrections. As a result, even if we add some new physics contributions to the theoretical predictions, the total branching ratio often agrees with the experimental value within those theoretical uncertainties. While tremendous efforts in order to improve the precision of the theoretical prediction have been made so as to match to the experimental precision, which could become even higher in the future machines, it is necessary to investigate the characteristics of the particles inside of the loop of  the $b\to s\gamma$ process using another kind of observables. In this article, we discuss a measurement of the circular-polarization of the photon of the $b\to s\gamma$ process, which  the left- and right-handedness of the couplings of the interactions among the particles inside of the loop. In the SM, the fact that the $W$ boson couples predominantly to the left-handed quarks induce the photon polarization to be mostly left-handed. On the other hand, many new physics  models contain new particles which couple differently from the SM. Therefore, the measurement of the photon polarization can be an useful tool to distinguish the interactions of the particles inside of the $b\to s\gamma$ loop from the SM-like one. 

Although, there have been several proposals for how to measure this photon polarization, its precise measurement has not been achieved yet. In this paper, we revisit the method proposed by Gronau et al.~\cite{Gronau} (the GGPR method in the following) using the exclusive $B\to K_{\rm resonance}\gamma$ followed by the three body decay of the $K_{\rm resonance}$.  Most interestingly, the Belle collaboration recently observed one of these decay channels, $B\to K_1(1270)\gamma\to(K\pi\pi)\gamma$, and found a relatively large branching ratio: $Br(B^+\to K_1^+(1270)\gamma )=(4.3\pm0.9(\text{stat})\pm0.9(\text{syst}))\times10^{-5}$~\cite{Belle-BK1gamma}, which dominates over the decay to $K_1(1400)$, previously studied in detail by GGPR~\cite{Gronau}. Thus, it is interesting to reconsider the feasibility of this method. In this article, we introduce a new  variable, $\omega$, which is originally proposed by Davier et al~\cite{DDLR} for the $\tau$ polarization measurement at LEP (the DDLR method in the following). As we show later-on, the fact that the decay width of $\threebody$ process depends only linearly on the polarization parameter $\lamg$ allows us to use the variable $\omega$ in our study. And the simplification of the fit by using $\omega$ makes it easier to include to the fit not only the angular dependence of the polarization parameter but also the three body Dalitz variable dependence, which improves the sensitivity to the polarization parameter as also pointed out in~\cite{DDLR}. On the other hand, the new radiative decay, to $K_1(1270)\gamma$ instead of the $K_1(1400)\gamma$, implies a more complex pattern of hadronic decay channels, not only through $K^*\pi$, but also through $K\rho$ and a possible $\kappa\pi$. In this work, we discuss, in details, the hadronic parameters required in this analysis. In particular, having various difficulties to extract them fully from the currently available experimental data, we attempt to evaluate them with a help of the so-called $^3P_0$ decay model. 

In section 2, we show a demonstration of the photon of the $b\to s\gamma$ being predominantly left-handed in the SM. We also discuss briefly  the contamination from the right-handed polarization.  
In section 3, we derive the master formula for the decay width of the $\threebody$ and the hadronic parameters needed in this formula are evaluated in section 4.  
In section 5, we introduce the new variable $\omega$. We show our numerical results in section 6, including the comparison of the sensitivity of the DDLR method with the other proposed methods . Section 7 is our conclusions.  

%
\section{Photon polarization of the $b\to s\gamma$ in SM}\label{sec:2}
%
In SM, the quark level $b\to s\gamma$ vertex without any QCD correction is given as:
\be\label{eq:1}
\overline{s}\Gamma (b\to s\gamma)_\mu b = \frac{e}{(4\pi)^2}\frac{g^2}{2M_W^2} V_{ts}^*V_{tb}F_2
\overline{s} i\sigma_{\mu\nu} q^\nu \left(m_b\frac{1+\gamma_5}{2}+m_s \frac{1-\gamma_5}{2}\right)b
\ee
where $q=p_b-p_s$ with $p_b$ and $p_s$ four-momentum of $b$ and $s$ quark, respectively, $F_2$ is the loop function, whose expression can found in~\cite{InamiLim}. When we fix the three momentum direction, namely the $q$ direction as $+z$ in the $b$ quark rest frame, one can compute explicitly the helicity amplitude and we readily find that the first (second) term is non-zero only when we multiply the left (right)-handed circular-polarization vector, which is defined as: 
\be
\epsilon_L^{\mu}=\frac{1}{\sqrt{2}}(0,1,-i,0), \quad
\epsilon_R^{\mu}=\frac{1}{\sqrt{2}}(0,1,i,0) 
\label{eq:2}
\ee 
Since $m_s/m_b\simeq 0.02 \ll 1$, the photon in $b\to s\gamma$ in SM is known to be predominantly left-handed. 

Once we include the QCD corrections, the other types of Dirac structure contribute and the above conclusion can be slightly modified. The result can typically described in term of the following effective Hamiltonian: 
\begin{equation}\label{eq:Heff}
\mathcal{H}_{eff}=-\frac{4G_F}{\sqrt 2}V_{ts}^*V_{tb}\left(\sum_{i=1}^6 C_i(\mu)\mathcal{O}_i(\mu)+C_{7\gamma}(\mu)\mathcal{O}_{7\gamma}(\mu)+C_{8g}(\mu)\mathcal{O}_{8g}(\mu)\right)
\end{equation}
where $C_i$ are the short-distance Wilson coefficients that can be calculated in perturbation theory and $\mathcal{O}_i$ are the local four-quark operators ($i=1\dots6$) and $\mathcal{O}_{7\gamma}$ and $\mathcal{O}_{8g}$ are the electro-magnetic and  chromo-magnetic penguin operators, respectively. The $\mu$ is the renormalization scale which is chosen as around $m_b$. 
Note that $\mathcal{O}_{7\gamma} = \frac{e}{16\pi^2}m_b\bar{s}_{L\alpha}\sigma_{\mu\nu}F^{\mu\nu}b_{R\alpha} $ is equivalent to the first term in Eq. (\ref{eq:1})\footnote{The term proportional to $m_s$ (the second term in Eq. (\ref{eq:1})) is neglected in this expression due to its smallness. }. 
%
%
In addition to the small $m_s/m_b$ contribution, there is potentially non-negligible right-handed pollution due to the perturbative and non-perturbative contributions. A numerical estimate for them is extremely important while it is currently not available for $B\to K_1\gamma$. On the other hand, many efforts have been made in the case of $B\to K^*\gamma$ using various QCD-based approaches~\cite{Khodjamirian}-\cite{Ball}. Note that the time-dependent CP asymmetry of $B\to K^*\gamma$ can also be used to determine the photon polarization (see Section 6 for some discussions). To have an idea, it is found that the most recent estimate for $B\to K^*\gamma$~\cite{Ball} shows that the right-handed correction is less than 1\%, while another estimate~\cite{Grinstein} shows that it can be up to 10\%.

On the other hand, when we consider the new physics contributions, the right-handed contribution can be significantly enlarged from different types of Dirac structure that those new physics models can induce. It should be emphasized that there are many of the new physics models which can accommodate  e.g. a large coefficient to the right-handed electro-magnetic operator ($O_{7\gamma}$ with the subscripts $L$ and $R$ interchanged) {\it without contradicting to the precise measurement of the inclusive $B\to X_s\gamma$ branching ratio}, for example Everett et al.~\cite{Everett}. 
%
\section{{The $B\to K_1\gamma \to (K\pi\pi)\gamma$ decays}}
\label{sec:3}
\subsection{{Master formula for $B\to K_1\gamma \to (P_1P_2P_3)\gamma$ decays}}
\label{sec:3}
Due to the angular momentum conservation and the fact that $B$-meson is a pseudo-scalar meson, helicity is conserved. Thus in order to determine the photon polarization it is sufficient to measure the polarization of the axial-vector meson $K_1(1^+)$ through its three body decay. 
As the physical final state $K_1$ must have either left- or right-handed polarization, the decay width can be written as 
\be
\Gamma(\Bbar\to \Kbar_1\gamma )=
\Gamma(\Bbar\to \Kbar_{1L} \gamma_L)+
\Gamma(\Bbar\to \Kbar_{1R} \gamma_R)
\ee
If we assume the narrow width of $K_1$, one can write the total quasi-four body decay width by these two terms, respectively, followed by the three body decay widths   
\[\Gamma(\Kbar_{1L}\to P_1 P_2 P_3), \quad \Gamma(\Kbar_{1R}\to P_1 P_2 P_3)\]

\begin{figure}[h]
\begin{center}
\epsfig{file=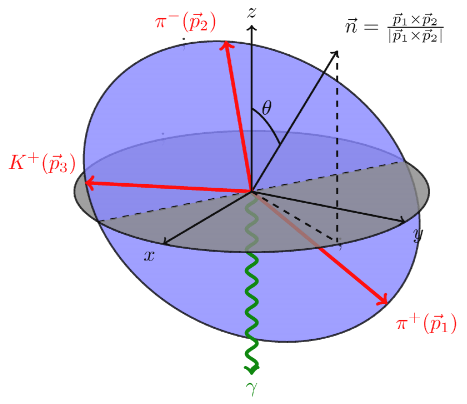,width=0.5\linewidth}
\caption{The $K_1\to K\pi\pi$ decay plane in the  rest frame of $K_1$. 
Defining the $-z$ direction as the photon direction, the $\theta$ is given as 
$\cos\theta\equiv\left(\frac{\vec{p}_{1}\times \vec{p}_{2}}{|\vec{p}_{1} \times \vec{p}_{2}|}\right)_z$.}
\label{fig:1}
\end{center}
\end{figure}

However, the width of the $K_1$'s is not really negligible ($\Gamma(K_1(1270))=90$~MeV, $\Gamma(K_1(1400))=174$~MeV according to PDG).  Therefore, we present for completeness, in the following, a prescription that includes the initial state width of the $K_1$ decay into the three body final state assuming the Breit-Wigner form, but which will not be used in practice. The Breit-Wigner factor is common to both polarizations and appears in modulus squared (therefore, its phase does not affect the crucial interference between the $\jay$ and $\jay^{*}$  terms below). Thus, our decay widths can be written as:
\be
\begin{split}
\frac{d\Gamma(\Bbar\to \Kbar_{1} \gamma\to (P_1P_2P_3)\gamma)}{dsds_{13} ds_{23}d\cos\theta} & \propto\sum_{{\rm pol.}=L, R}\Gamma(\Bbar\to \Kbar_{1{\rm pol.}} \gamma_{\rm pol.}) \\ 
& \times\frac{d\Gamma(\Kbar_{1{\rm pol.}}\to P_1 P_2 P_3)}{ds ds_{13} ds_{23}d\cos\theta} \times\frac{1}{(s-m_{K_1}^2)^2+m_{K_1}^2 \Gamma_{K_1}^2}
\end{split}
\label{eq:5}
\ee
where 
$s=(p_{1}+p_{2}+p_{3})^2$ is the off-shell "$p^2$" of the $K_1$ and $s_{ij}=(p_{i}+p_{j})^2$ with $p_i$ to be the four-momentum of the final state $P_i$. 
Defining the $-z$ direction as the photon direction in the $K_1$ rest frame (see Fig.~\ref{fig:1}), the $\theta$ is given as $\cos\theta\equiv\left(\frac{\vec{p}_{1}\times \vec{p}_{2}}{|\vec{p}_{1} \times \vec{p}_{2}|}\right)_z$. 

The kinematic distribution of this three body decays carries the information of the $\Kbar_1$ polarization. It is important to notice that the polarization information we would like to obtain is the difference between $\Gamma(\Bbar\to \Kbar_{1L} \gamma_L)$ and $\Gamma(\Bbar\to \Kbar_{1R} \gamma_R)$ in Eq. (\ref{eq:5}) while experimentally, only the L.H.S of this formula, i.e. the total decay width can be measured. Thus, the high sensitivity to the polarization information can be achieved only if there is  a significant difference in the decay distributions between $\Kbar_{1L}$ and $\Kbar_{1R}$. 

The differential decay width of $\Kbar_{1L,R}$ decay can be described by the helicity amplitude, $\jay_\mu$, which we define as: 
\be
\mathcal{M}(\Kbar_{1L,R}\to P_1P_2P_3) = \epsilon_{K_1 L,R}^{\mu} \jay_\mu.
\label{eq:6}
\ee
Considering that $\jay_{\mu}$ represents the decay amplitude of the $K_1$ decaying into three pseudoscalar mesons, we  can parameterize it in terms of two functions $\see_{1,2}$: 
\be
\jay_\mu=\see_1(s, s_{13},s_{23})p_{1\mu}-\see_2(s, s_{13},s_{23})p_{2\mu}
\label{eq:7}
\ee
where we omitted to write explicitly the Dalitz and angular variable dependences of $\jay_\mu$.  Note the $s$ dependence of the coefficients, which means that in principle there could be some dependence on the off-shell $p^2$ of the $K_1$. Nevertheless, this dependence is not important as soon as the integration is limited to the $K_1$ bump, especially for the ratio $\omega$ which is the relevant quantity in our method (see next section).
The detailed expressions of $\see_{1,2}(s, s_{13},s_{23})$ for given channels are derived in the next section but here we note that $\see_{1,2}(s, s_{13},s_{23})$ can contain complex numbers. 
Using the definition of the helicity in Eq.~(\ref{eq:2}), one can easily find in the $K_1$ reference frame
\be
\label{Eq:8}
\begin{split}
 \frac{d\Gamma(\Kbar_{1L,R}\to P_1 P_2 P_3)}{ds ds_{13} ds_{23}d\cos\theta} &\propto|\mathcal{M}(\Kbar_{1L,R}\to P_1P_2P_3)|^2 \\
&\propto \frac{1}{4}|\vec{\jay}|^2(1+\cos^2\theta)\mp \frac{1}{2} Im [\vec{n}\cdot (\vec{\jay}\times \vec{\jay}^*)]\cos\theta
\end{split}
\ee
where $\vec{n}\equiv \frac{\vec{p}_1\times\vec{p}_2}{|\vec{p}_1\times\vec{p}_2|}$ so that: 
\bea
|\vec{\jay}|^2 &=& |\see_{1}|^2 |\vec{p}_1|^2+|\see_{2}|^2 |\vec{p}_2|^2-(\see_{1}\see_{2}^*+\see_{1}^*
\see_{2})(\vec{p}_1\cdot \vec{p}_2) \\
\vec{n}\cdot (\vec{\jay}\times \vec{\jay}^*)&=& -(\see_{1}\see_{2}^*-\see_{1}^*\see_{2})|\vec{p}_1\times \vec{p}_2|
\eea 
where $\vec{p}_i\cdot \vec{p}_j=E_iE_j-(s_{ij}-m_i^2-m_j^2)/2$ and $|\vec{p}_i\times  \vec{p}_j|=\vec{p}_i\cdot \vec{p}_j \tan^{-1}\phi$  
with $E_i=(s-s_{jk}+m_i^2)/(s\sqrt{s})$ and $\phi=\cos^{-1}[(\vec{p}_i\cdot \vec{p}_j)/(|\vec{p}_i||\vec{p}_j)]$. 

It is worth mentioning that the difference between the left- and right-handed polarization amplitudes comes from the second term of Eq. (\ref{Eq:8}) which, to be non vanishing, requires the amplitude $\jay$ to contain more than one amplitude with non-vanishing relative phase. 
Such a condition can be nicely realised in this decay channel since when $K_1$ decays into three body final states through more than one intermediate two-body channels, such as $K^*\pi$ and $K\rho$, there is the non-vanishing relative phase originated from their Breit-Wigner forms (based on the isobar model).  

Finally, the master formula is obtained in terms of the polarization parameter $\lamg$ (the $B\to K_1$ form factor is skipped as being a common factor)
\be
\begin{split}
\frac{d\Gamma(\Bbar\to \Kbar_{1} \gamma\to (P_1P_2P_3)\gamma)}{dsds_{13} ds_{23}d\cos\theta}  &\propto \frac{1}{(s-m_{K_1}^2)^2+m_{K_1}^2 \Gamma_{K_1}^2} \\
 &\times \left\{\frac{1}{4}|\vec{\jay}|^2(1+\cos^2\theta)+\lamg \frac{1}{2} Im [\vec{n}\cdot (\vec{\jay}\times \vec{\jay}^*)]\cos\theta\right\}
 \end{split}
 \label{eq:9}
\ee
with 
\be
\lamg \equiv \frac{\Gamma(\Bbar\to \Kbar_{1R} \gamma_R)-\Gamma(\Bbar\to \Kbar_{1L} \gamma_L)}{\Gamma(\Bbar\to \Kbar_{1} \gamma)}
\ee
which agrees with the expression in~\cite{Gronau}.

Since the two $K_1$ resonances, $K_1(1270)$ and $K_1(1400)$,   are rather close each other, one could expect interference between them, which should then be taken into account in the formulas. 
On the other hand, the Belle data~\cite{Belle-BK1gamma} shows a suppression of the $K_1(1400)$. In~\cite{Hatanaka,Lee} it has been shown that such a suppression can be explained by taking into account the fact that these two states are the mixture of $1^3P_1$ and $1^1P_1$ sates and a reasonable choice of the mixing angle can explain such a suppression (see section 4 for more detailed discussion on the mixing angle).

\subsection{The $\see_{1,2}$ functions for the $K_1\to K\pi\pi$ decays} 
\label{sec:4}
In this section, we derive the $\see_{1,2}$ function, which is defined in Eqs. (\ref{eq:6}) and (\ref{eq:7}) for the $K_1(1270/1400)$ decay. 
The three body decay channels of the $K_1(1270/1400)$ are the $\pi\pi K$ final states.  
We first assume that this three pseudoscalar meson final state comes form the quasi-two-body decay through a vector meson, namely $\rho$ or $K^*$. 
The different decay channels and the possible vector resonances for $K_1^+(1270/1400)$ and $K_1^0(1270/1400)$ are listed below. 
\bea
I: \quad &K_1^+(1270/1400)& \to \pi\underbracket{\underbracket{^0(p_1)\pi}_{\rho^+}\overbracket{\phantom{}^+(p_2)K}^{K^{*+}}}_{K^{*0}}\phantom{}^0(p_3)
\label{eq:14}\\
II: \quad &K_1^+(1270/1400)& \to \pi\underbracket{\underbracket{^-(p_1)\pi}_{\rho^0}\phantom{}^+(p_2)K}_{K^{*0}}\phantom{}^+(p_3)
\label{eq:15}\\
III: \quad &K_1^0(1270/1400)& \to \pi\underbracket{\underbracket{^0(p_1)\pi}_{\rho^-}\overbracket{\phantom{}^-(p_2)K}^{K^{*0}}}_{K^{*+}}\phantom{}^+(p_3)
\label{eq:16}\\
IV: \quad &K_1^0(1270/1400)& \to \pi\underbracket{\underbracket{^+(p_1)\pi}_{\rho^0}\phantom{}^-(p_2)K}_{K^{*+}}\phantom{}^0(p_3)
\label{eq:17}
\eea
The decay amplitudes for these decay channels can be written as the sum of the amplitude with different intermediate vector meson channel: 
\be
\mathcal{M}(K_1\to P_1P_2P_3)
=\sum_V
c_{ijk} \mathcal{M}_{(P_iP_j)P_k}^V
\ee
where $P_{1,2,3}$ represent the final state mesons carrying the momentum $p_{1,2,3}$ as assigned in the Eqs. (\ref{eq:14}) to (\ref{eq:17}) and $V$ represents the vector meson resonance.  
The Clebsch-Gordan coefficients, $c_{ijk}$,   for each intermediate channel are given as: 
\bea
\mathcal{M}_I(K_1^+\to \pi^0(p_1)\pi^+(p_2)K^0(p_3))&=&
\frac{\sqrt{2}}{3}\mathcal{M}_{(P_1P_3)P_2}^{K^{*0}}
-\frac{\sqrt{2}}{3}\mathcal{M}_{(P_2P_3)P_1}^{K^{*+}}
+\frac{1}{\sqrt3}\mathcal{M}_{(P_1P_2)P_3}^{\rho^{+}} \no
\mathcal{M}_{II}(K_1^+\to \pi^-(p_1)\pi^+(p_2)K^+(p_3))&=&
-\frac{{2}}{3}\mathcal{M}_{(P_1P_3)P_2}^{K^{*0}}
-\frac{{1}}{\sqrt{6}}\mathcal{M}_{(P_1P_2)P_3}^{\rho^{0}} \no
\mathcal{M}_{III}(K_1^+\to \pi^0(p_1)\pi^-(p_2)K^+(p_3))&=&
\frac{\sqrt{2}}{3}\mathcal{M}_{(P_1P_3)P_2}^{K^{*+}}
-\frac{\sqrt{2}}{3}\mathcal{M}_{(P_2P_3)P_1}^{K^{*0}}
+\frac{1}{\sqrt3}\mathcal{M}_{(P_1P_2)P_3}^{\rho^{-}} \no
\mathcal{M}_{IV}(K_1^+\to \pi^+(p_1)\pi^-(p_2)K^0(p_3))&=&
-\frac{{2}}{3}\mathcal{M}_{(P_1P_3)P_2}^{K^{*+}}
-\frac{{1}}{\sqrt{6}}\mathcal{M}_{(P_1P_2)P_3}^{\rho^{0}} \nonumber 
\eea

Using the detailed expression for the quasi-two-body decay amplitude $\mathcal{M}_{(P_iP_j)P_k}^V$  given in Appendix A, we find: 
\bea
&&\mathcal{M}(K_{1L,R} \to \pi\pi K)_{A=I\sim IV}  =
\epsilon_{K_1 L,R}^{\mu} \jay_\mu^A \\
&& \jay_\mu^A=\see_1^A(s, s_{13},s_{23})p_{1\mu}-\see_2^A(s, s_{13},s_{23})p_{2\mu}
\eea
with 
\be
\begin{split}
\see_1^{I,III} &=
\frac{\sqrt{2}}{3}(a_{13}^{K^*}-b_{13}^{K^*})
+\frac{\sqrt{2}}{3}b_{23}^{K^*}
+\frac{1}{\sqrt3}a_{12}^{\rho}, \quad  
\see_1^{II,IV}=
-\frac{2}{3}(a_{13}^{K^*}-b_{13}^{K^*})
-\frac{1}{\sqrt{6}}a_{12}^{\rho} \\
\see_2^{I,III} &=
\frac{\sqrt{2}}{3}b_{13}^{K^*}
+\frac{\sqrt{2}}{3}(a_{23}^{K^*}-b_{23}^{K^*})
-\frac{1}{\sqrt3}b_{12}^{\rho}, \quad
\see_2^{II, IV} =
-\frac{2}{3}b_{13}^{K^*}
+\frac{1}{\sqrt{6}}b_{12}^{\rho} 
\end{split}
\ee
where 
\be
\begin{split}
a_{ij}^V &= g_{VP_iP_j}{\rm BW}_V(s_{ij})[f_V+h_V\sqrt{s}(E_i-E_j)-\Delta_{ij}] \\
b_{ij}^V &= g_{VP_iP_j}{\rm BW}_V(s_{ij})[-f_V+h_V\sqrt{s}(E_i-E_j)-\Delta_{ij}]
\end{split}
\ee
with $\Delta_{ij}\equiv \frac{(m_i^2-m_j^2)}{M_{ij}^2}[f+h\sqrt{s}(E_i+E_j)]$. 
%

\section{{Hadronic parameters and their estimation in the $^3P_0$ model}} 
\label{sec:6}

The next step is to obtain the coupling constants and the form factors determining the above functions $\see_{1,2}$, i.e. the following hadronic parameters
\[
g_{\rho\pi\pi}, \quad g_{K^*K\pi}, \quad f_V, \quad h_V
\]
Noting that there are total of four $f_V$ and $h_V$ ($V=\rho, K^*$) for each $K_1(1270)$ and $K_1(1400)$, we have ten free parameters in this decay mode. One may consider the relative phases between the form factors $f_V$ and $h_V$, which increases the number of free parameter. However, this phase could actually be determined  theoretically or experimentally.

Ideally, these parameters should be extracted from the same experimental data of the $B\to K_1\gamma$ decay, which allows us a hadronization model independent analysis. For the present moment, one can use the other experimental information such as $\rho\to\pi\pi$ and $K^*\to K\pi$ for the $g$ couplings and  
$K_1\to K\rho$ and $K_1\to K^*\pi$ for $f_V$ and $h_V$ form factors.   
We first present in the following subsection how to relate those experimental information to our hadronic parameters.

\subsection{Hadronic parameters}
\underline{\it The $V PP$ coupling constant: $g_{VPP}$:}\\
The $g_{VPP}$ coupling can be extracted from the partial decay width of the vector mesons. These are well measured for $V=\rho, K^*$ so that we can obtain this coupling rather precisely. The partial decay width can be written as: 
\be
\Gamma (V \to P_1P_2) =\frac{g_{VP_1P_2}^2}{2\pi m_V^2} |\vec{p}|^3 \frac{1}{3}
\ee
where $|\vec{p}|=\sqrt{(m_V^2-(m_1+m_2)^2)(m_V^2-(m_1-m_2)^2)}/2m_V$. 
Then, using the experimental values of $\rho$ and $K^*$ widths, we find
\be
g_{\rho\pi\pi}=(5.98\pm0.02), \quad g_{K^*K\pi}=(5.68\pm0.05)
\ee
\bigskip

\noindent
\underline{\it The $K_1\to VP$ form factors $f_V$ and $h_V$}\\
To describe the $K_1\to VP$ decay, we used two independent form factors $f_V$ and $h_V$
\begin{equation}
	\langle V(p_{V},\varepsilon^{(V)}) P_k(p_k))|\Delta H_A|K_1(p_{K_1},\varepsilon^{(_{K_1})})\rangle=\varepsilon_\mu^{({K_1})}T^{\mu\nu}\varepsilon_\nu^{(V)*}~,~~T^{\mu\nu}=f_Vg^{\mu\nu}+h_V p_{V}^{\mu}p_{K_1}^{\nu}.
	\label{eq:AVP1}
\end{equation}
On the other hand, the $K_1\to VP$ can also  be written in terms of the helicity amplitudes for the two possible $+z$ spin projection of $K_1$ and the vector meson, 
$(\lambda_{K1}, \lambda_{V})=(0,0)$ and $(1,1)$.  These two amplitudes actually can be written in terms of common partial wave amplitudes. Thus, when we expand them up to $L=2$, we can equivalently write these helicity amplitudes by the two partial wave amplitudes~\cite{Chung}:
\be
\langle V(-\vec{p}_k,\lambda_{V})P_k(\vec{p}_k)|\Delta H_A|K_1(\vec 0,\lambda_{K_1})\rangle = (A_V^S+\sqrt5\langle2,0;1,\lambda_V|1,\lambda_V\rangle A_V^D) D_{\lambda_{K_1},\lambda_V}^{1*}(\Omega_V)
\label{eq:PWA}
\ee
where $A_V^{S,D}$ are the partial wave amplitudes. Then, these amplitudes can be 
experimentally extracted through the partial wave analysis of the $K_1\to VP$ processes using: 
\be
\begin{split}
\Gamma(K_1\to VP)_{S-{\rm wave}}&= \frac{|\vec{p}_V|}{8\pi s_A}|A^S_V|^2\\
\Gamma(K_1\to VP)_{D-{\rm wave}}&= \frac{|\vec{p}_V|}{8\pi s_A}|A^D_V|^2
\end{split}
\ee
Comparing to Eq. ~\eqref{eq:AVP1} and Eq.~\eqref{eq:PWA}, we can immediately find the relation between the two form factors and the partial wave amplitudes ($f_V$, $h_V$ depend in general on $E_V$: 
\be
\begin{split}
f_V &= -A^S_V-\frac{1}{\sqrt2}A^D_V \\
h_V &= \frac{E_V}{\sqrt{s_{K_1}}|\vec{p}_V|^2}\left[\left(1-\frac{\sqrt{s_V}}{E_V}\right)A^S_V+\left(1+2\frac{\sqrt{s_V}}{E_V}\right)\frac{1}{\sqrt2}A^D_V\right]
\end{split}
\ee

\vskip 1cm
Partial wave analysis of $K_1\to VP$ process has indeed been performed by the ACCMOR collaboration~\cite{Daum} and a very precious information related to $K_1$ meson has been extracted, which constitute the basis of the PDG entries. It is the currently available most extensive study of the $K\pi\pi$ channels, with full angular distributions analysis, determination of relative phases between all amplitudes.   On the other hand, the interpretation of the ACCMOR data contains various problems in the theoretical point of view, or even empirically 
\footnote{We give some examples of those problems:
\begin{itemize}
\item The resonance study is done by using the K-matrix method. Therefore, to match the information obtained from their analysis to our Breit-Wigner parametrisation is not a simple task. Masses and widths must be recalculated. In addition, the authors use a complex phase space. 

\item ACCMOR results are obtained by using particular models for strong interaction  production through $K p$ scattering, like the Deck effect. Moreover, and more worrying, the $D$ wave in $K_1(1270) \to K^*\pi$  is depending strongly on the production transfer $t$. This fact may escape attention of PDG readers, because it averages between the two sets of data (high $t$, low $t$). 

\item There are only limited amount of data on the $D$-waves. They are poorly measured in $K^*\pi$. For the D-wave amplitude $K_1(1270)\to K\rho$, there is no information...

\end{itemize}. 
Two other items are found to be important issues for the determination of polarization, the question of relative phases between the various partial waves, and the $\kappa \pi$ channel; we devote to them two separate paragraphs below. It is to  be noted that the Babar collaboration~\cite{Babar-BK1pi} has performed a reanalysis of the ACCMOR data ; it contains useful complementary information, with somewhat different results for the parameters. On the other hand, ref.~\cite{Belle-BpsiK1} is a new, completely independent, analysis, which comes to certain conclusions differing from ACCMOR, especially for the $\kappa\pi$ channel.\vskip 0.5 cm}. We will come back to some of these issues later in this section. In any case, we found that it is currently impossible to extract all the parameters from experimental data. Thus, we need a help of theoretical model inputs for this reason. In the following, we try to use the so-called  $^3P_0$ model, that is an intuitive model describing the decay by the creation of a quark-antiquark pair.

\subsection{Estimating the hadronic parameters in the $^3P_0$ model.} 
The $^3P_0$ model \footnote{This model has been first developed in~\cite{LOPR} and then extensively discussed by the group around N.~Isgur~\cite{Kokoski,Godfrey,Blundell1} in Canada. It has been already used by Blundell et al. in the present context~\cite{Blundell2}.} has the advantage that it provides rather complete predictions-in particular the model fixes the coupling signs, ratios of $K^* \pi$ to $K \rho$ couplings, $D/S$ ratios and the full set of couplings once the quark pair-creation constant $\gamma$ is fixed. The model fixes the ratio of two independent couplings ($C=\pm1$) which is left free by the $SU(3)$ symmetry. Another illustrative example of its specific usefulness is the prediction of a very small decay of $K_1(1270)$ into $K_0^*(1430)$. On the other hand, we must stress that it is a very approximate model, not claiming to be always quantitative. Its main drawback is that it is essentially non relativistic.

There are two independent $K_1$ states in the quark model, the $1^3P_1$ and $1^1P_1$ states, which are called $K_{1A}$ and $K_{1B}$ respectively. With the $^3P_0$ model, we can predict the decay rates of these two states. However, these are not the physical mass eigenstates $K_1(1270, 1400)$: it has been known that the observed hierarchy of decays into $K^*\pi$ and $K \rho$, can be nicely explained by considering that the physical states are a mixture of $K_{1A}$ and $K_{1B}$ with a mixing angle $\theta_{K_1}$: 
\be
\begin{split}
|K_1(1270)\rangle &= |K_{1A}\rangle\sin\theta_{K_1}+|K_{1B}\rangle\cos\theta_{K_1} \\
|K_1(1400)\rangle &= |K_{1A}\rangle\cos\theta_{K_1}-|K_{1B}\rangle\sin\theta_{K_1}
\end{split}
\label{eq:K1mixing}
\ee
Then, the 8 independent amplitude ($A^{S}_{K^*/\rho}, A^{D}_{K^*/\rho}$) each for $K_1(1270)$ and  $K_1(1400)$ can be reduced to the 4 amplitudes $A^{S/D}_{K^*/\rho}$ and one mixing angle $\theta_{K_1}$:
\be
\begin{split}
	A^S_{K_1(1270)\to K^*\pi/K\rho} &= S_{K^*/\rho}(\sqrt{2}\sin\theta_{K_1}\mp\cos\theta_{K_1}) \\
	A^D_{K_1(1270)\to K^*\pi/K\rho} &= D_{K^*/\rho}(-\sin\theta_{K_1}\mp\sqrt{2}\cos\theta_{K_1}) \\
	A^S_{K_1(1400)\to K^*\pi/K\rho} &= S_{K^*/\rho}(\sqrt{2}\cos\theta_{K_1}\pm\sin\theta_{K_1}) \\
	A^D_{K_1(1400)\to K^*\pi/K\rho} &= D_{K^*/\rho}(-\cos\theta_{K_1}\pm\sqrt{2}\sin\theta_{K_1})
\end{split}
\label{SDamplitudes}
\ee
The $S_{K^*/\rho}$, $D_{K^*/\rho}$ amplitudes are expressed in terms of the hadron wave functions and of the quark-pair creation constant $\gamma$ in this model. 
Having these model parameters fixed, we can obtain the mixing angle by fitting to the experimental data of $K_1$ decays. 
The available experimental information is listed below: 
\be
\begin{split}
Br(K_1(1270)\to K^*\pi):\ Br(K_1(1270)\to K\rho) &= (16\pm 5)\% :\  (42\pm 6)\%  \\
Br(K_1(1400)\to K^*\pi):\ Br(K_1(1400)\to K\rho) &= (94\pm 6)\% :\  (3.0\pm 3.0)\%  \\
\frac{Br(K_1(1400)\to K^*\pi)_{D-{\rm wave}}}{Br(K_1(1400)\to K^*\pi)_{S-{\rm wave}}} &=  0.04 \pm 0.01 \\
\frac{Br(K_1(1270)\to K^*\pi)_{D-{\rm wave}}}{Br(K_1(1270)\to K^*\pi)_{S-{\rm wave}}} &= 0.54 \pm 0.15~(\text{low}~t) \\
\frac{Br(K_1(1270)\to K^*\pi)_{D-{\rm wave}}}{Br(K_1(1270)\to K^*\pi)_{S-{\rm wave}}} &= 2.67 \pm 0.95~(\text{high}~t) 
\end{split}
\ee
where $t$ is the resonance production momentum transfer. 
We will present the details of the fitting procedure and the estimate of the theoretical uncertainties in the forthcoming paper~\cite{TKL}. Here, we give, for indication, the result with the following conditions: i) we adopt a set of harmonic oscillator wave functions with a common h.o. radius $R=1/(0.4~\text{GeV})=2.5~$GeV$^{-1}$ ($\psi(r) \propto exp(-r^2/2R^2)$), which, we would say, is a widespread and empirically satisfactory recipe ii) we use $\gamma\sim 4$ iii) we consider only low~$t$ data for the $K_1(1270)$ decay, iv) we assume all the particles are on shell, except for the $K_1(1270) \to K \rho$, v) we use the damping factor to be $\beta'=3$~GeV$^{-2}$ (see the discussion on the damping factor in Appendix~B).  As a result, we find that the mixing angle $50^{\circ}-60^{\circ}$  is  well compatible with the data. Then, the partial wave amplitudes can be given, for example for $\theta_K=60^{\circ}$, as  
\be
\begin{split}
A_{K_1(1270)\to K^*\pi}^S  &\sim 1.6, ~ ~\quad A_{K_1(1270)\to K^*\pi}^D  \sim -0.2 \\ 
A_{K_1(1400)\to K^*\pi}^S  &\sim 3.1, ~ ~\quad A_{K_1(1400)\to K^*\pi}^D  \sim 0.2 \\ 
A_{K_1(1270)\to K\rho}^S &\sim 4.6, ~ ~\quad A_{K_1(1270)\to K\rho}^D ~\sim -0.03 \\ 
A_{K_1(1400)\to K\rho}^S &\sim -0.5,\quad A_{K_1(1400)\to K\rho}^D ~\sim -0.4 
\end{split}
\ee

\subsection{Discussions on the phases and the scalar resonances }
\subsubsection{The relative phases between different $K_1 \to VP$ couplings}
The issue of the phases of the resonance couplings is very important in the present approach, since the dependence on $\lamg$ relies entirely on the phase of $\jay$. And, indeed, one finds that changing the relative sign of the decays of $K_1(1270)$ to $K\rho$ over $K^*\pi$ channel would entirely change the prediction for $\lamg$. Then, we formulate the following observations:

\begin{itemize}

\item Phases between all the various resonances and decay channels into $K\pi\pi$ are in principle measurable, and indeed are measured by ACCMOR collaboration. They can be also determined from our $^3P_0$ model. From a theoretical point of view, one must not forget that, to obtain the full coupling sign of a quasi two-body channel, one must take the product of the amplitude for the decay into the quasi two-body channel with the one of the isobar decay (e.g. $K^*\to K\pi$), in order to have the sign into the common final state $K\pi\pi$, the only to make sense in comparing channels.

However, the question remains far from trivial, because :

\item There were some misunderstandings in interpreting the ACCMOR data (e.g. the large D/S phase read by Gronau et al. does not in fact correspond to the $D/S$ relative phase for the couplings of the $1400$ ; the strong bump around 1400 in the $D$ wave phase diagram, Fig.~13 in ref.~\cite{Daum}, does not correspond to the $D$ wave of the $K_1(1400)$, which is very small ; it is a reflection of what happens in the $S$ wave, since $D$ wave phase is defined by reference to the S wave.).

\item Our model predicts real phases for all the couplings, which is also almost the case for the true $K-$matrix predictions of ACCMOR collaboration. On the other hand, the data of ACCMOR show something different; so-called off-set phases, imaginary, and not predicted by the true $K-$matrix, are to be added to describe the data . The origin of these additional phases is unknown. It is very important to realise that that the solid lines of the histograms in~\cite{Daum} do not represent the true $K-$matrix predictions, but include the ad hoc offset-phases. 

\item One can test the soundness of our model by checking whether the predicted relative signs of couplings to $K^*\pi$ and $K \rho$ agree with those shown by Daum et al.. In our study, we tried to establish the connection between the conventions of our model and those of ACCMOR collaboration. For the $A^D/A^S$ ratios in the common channel $K^*\pi$, the relation is trivial, and we find that there is agreement : $A^D/A^S<0$ for the $K_1(1270)$, $A^D/A^S>0$ for the $K_1(1400)$
\footnote{In the latter case, the ACCMOR sign is deduced from the reanalysis by Babar~\cite{Babar-BK1pi}}. On the other hand, for the relative sign between the amplitudes of $K_1(1270)\to(K^*\pi)_S$ and $K_1(1270)\to(K\rho)_S$, the conventions used by ACCMOR are not obvious, while this sign is crucial. In our study, we use the signs of $^3P_0$, but we also test different combinations of these relative signs, by allowing for additional phases $\delta$.   
\end{itemize}

\subsubsection{ The controversial $K_1(1270)\to K \pi \pi$ decay through a scalar meson: $K_1\to scalar+\pi$}
The PDG assigns a large branching ratio to this decay channel:
 $Br(K_1(1270)\to K_0^*(1430)\pi)=(28\pm4)\%$. It is  extracted as all the branching ratios, from the ACCMOR data and analysis. However, this interpretation has been questioned. The original ACCMOR measurement shows indeed a clear, strongly coupled peak in the (scalar + $\pi$) channel around the mass $1270$. However, it is not at all claimed that the scalar is $K_0^*(1430)$ ; it is treated as a lower and much broader scalar meson ($\Gamma\simeq600$~MeV ; or could be a continuum  $(K\pi)_{\rm{S-wave}}$ according to~\cite{Dunwoodie}. 
Indeed, our model predicts the decay to the $K_0^*(1430)\pi$ channel to be of the order of 1\%. 
Recently, the Belle collaboration has made a new branching measurement using the $B\to J/\psi (\psi^{\prime})K_1$ decay followed by $K_1\to K\pi\pi$. What is most striking is that indeed, Belle finds $Br(K_1(1270)\to K_0^*(1430)\pi)\simeq2\%$~\cite{Belle-BpsiK1}, as we predict, while not finding any new component in the $K_1$ decay: the $Br$ missing with respect to ACCMOR seems to be filled by an enlargement of $K\rho$. Therefore, in our analysis, we do not include the  $K_1(1270)\to K_0^*(1430)\pi$ channel. We do not include either any other possible scalar in the presented results. However, to take into account the contradictory conclusions of ACCMOR we have kept in mind the possibility that there is some significant portion of the branching ratio carried by a very wide scalar meson, different from the $K_0^*(1430)$, such as the low lying state $\kappa(800)$.

\section{Determination of $\lambda_\gamma$ in the DDLR method}
\label{sec:5}

In this section, we demonstrate how to determine the polarization parameter $\lamg$ from the actual experimental data  using the maximum likelihood method. In particular, we introduce the DDLR method which was first applied in the $\tau$ polarization measurement at the ALEPH experiment~\cite{DDLR}. 
In the maximum likelihood method, knowing the $\lamg$ dependence on the probability density function (PDF), the $\lamg$ closest to its true value can be obtained where the likelihood function  (or equivalently, log-likelihood) given by the $N$ sample of data takes its maximum value. In our case, the PDF, $W$, can be given as the decay width integrand normalized to unity (after due multiplication by the modulus squared of the Breit-Wigner). Let us reiterate our statement that when one remains within the bump of the $K_1$ resonance, the decay amplitude weakly depends on $s=p^2(K_1)$, and one can set $s=m_{K_1}^2$ in their expression, i.e. in the $\jay$'s, {\bf which we assume therefrom}.

Thus, using Eq. (\ref{eq:9}), we find
\be
W(s_{13}, s_{23}, \cos\theta) = f(s_{13}, s_{23}, \cos\theta) +\lamg g( s_{13}, s_{23}, \cos\theta)
\ee
where 
\be
\begin{split}
f(s_{13}, s_{23}, \cos\theta) &= \frac{1}{4I}|\vec{\jay}|^2(1+\cos^2\theta) \\
g(s_{13}, s_{23}, \cos\theta) &= \frac{1}{2I}Im [\vec{n}\cdot (\vec{\jay}\times \vec{\jay}^*)]\cos\theta \\
I &= \frac{2}{3} \int ds_{13}ds_{23} |\vec{\jay}|^2
\label{eq:fandg}
\end{split}
\ee
where $f$, $g$ are normalised relatively to the measure $ds_{13}ds_{23}d\cos\theta$.

Then, the likelihood function for the $N$ events of data can be given as
\be
\mathcal{L} = \prod_{i=1}^N \left[f(s_{13}^i, s_{23}^i, \cos\theta^i)+\lamg g(s_{13}^i, s_{23}^i, \cos\theta^i)\right]
\ee
where $i$ indicates the kinematic variable of each event. The true value of  $\lamg$ should maximize  this function, namely should be the solution of the following equation: 
\be
\frac{\partial {\mathcal{L}}}{\partial \lamg}=0
\ee

Now, we explain the DDLR method~\cite{DDLR}. The next procedure to look for the value of $\lamg$ in our problem is usually to use the known distribution of $f$ and $g$ functions and fit the value of $\lamg$ so as to maximize the likelihood function. It should be noted that this is not a very simple task, especially since the $f$ and $g$ are complicated functions as shown in Eq.~\eqref{eq:fandg} and the Appendix~A. In~\cite{DDLR}, it is pointed out that when the PDF depends on the parameter, which  we are interested in, {\it only linearly}, one can reduce such a multi-dimensional fit to an one-dimensional one using a single variable $\omega$  which is defined as follows: 
\be
\omega(s_{13}, s_{23}, \cos\theta) =\frac{g(s_{13}, s_{23}, \cos\theta)}{f(s_{13}, s_{23}, \cos\theta) }. 
\ee
This can be proved simply by writing down the log-likelihood of our problem: 
\be
\ln {\mathcal{L}}= \sum_{i=1}^N \ln \left[1+\lamg \omega(s_{13}^i, s_{23}^i, \cos\theta^i)\right] 
+ {\rm\ terms\ independent\ of\ } \lamg.
\ee
where $\lamg$ does not depend on $f$ and $g$ separately but only on their ratio $\omega$. 
This demonstrates that only the $\omega$ distribution is needed to extract $\lamg$. 

Another thing that is pointed out in~\cite{DDLR} is that the polarization parameter is often determined only by using the angular distribution, however, the sensitivity to it can be further improved by considering all the kinematic information, such as the Dalitz variable distribution. Therefore, we use the dependence of $\lamg$ not only on $\cos\theta$ but also on $s_{13}$ and $s_{23}$ in this work. 
Considering the fact that $f$, $g$ and $\omega$ have very complicated dependences on these kinematic variables, the reduction to the one-dimensional fit achieved by using the variable $\omega$ is very efficient for the data analysis as shown in the following. 

It must be underlined that, in the present case, in contrast with the initial DDLR problem, $\tau\to\pi\nu$, $\omega$ is not a purely kinematic variable, it depends itself on the theoretical model, as it is the case for $\tau \to a_1\nu$. Then the method to obtain the distribution in $\omega$ is as follows. First, following the standard MC method, we generate the faked events according to the PDF. Then, we compute the $\omega$ value for each event. In this way, we obtain the omega distribution $N^{\rm MC}(\omega)$ according to the PDF.

\begin{figure}[t]
\begin{center}
\epsfig{file=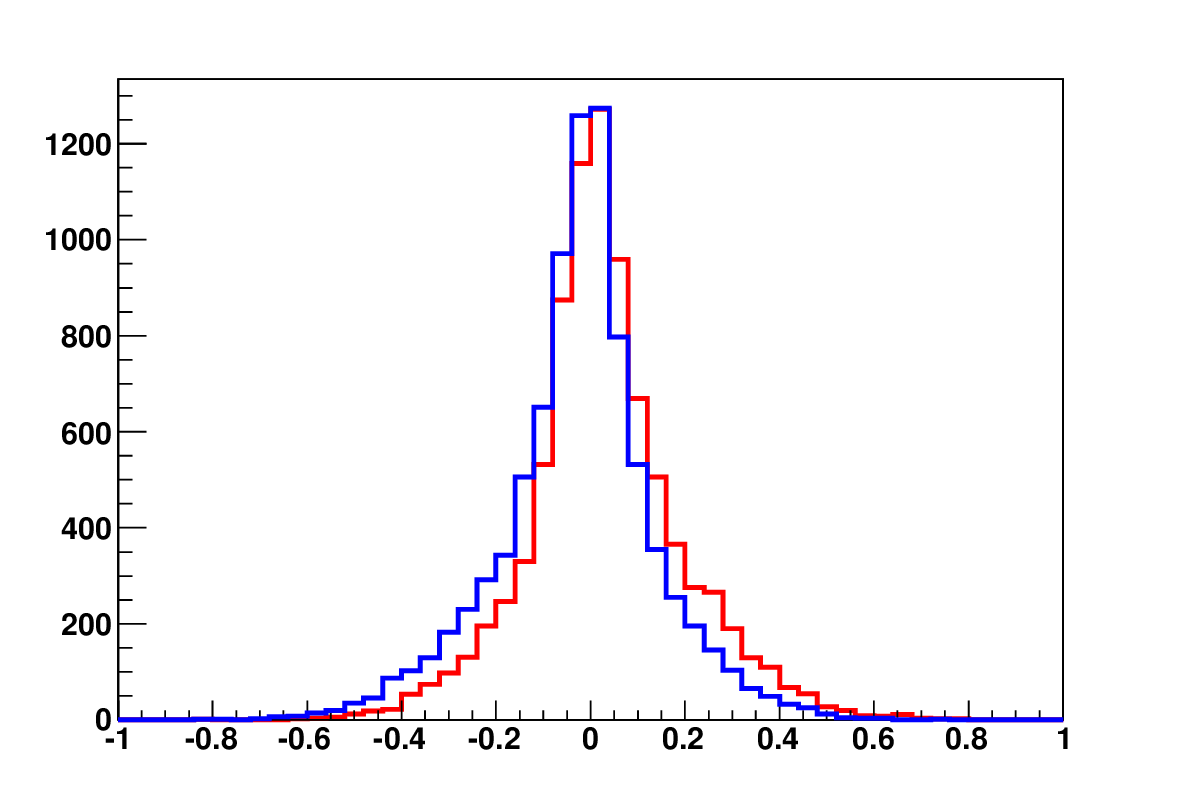,width=10.cm}
\caption{The simulated $\omega$-distribution for $\lambda_\gamma=+1$ (red) and $\lambda_\gamma=-1$ (blue). The polarization parameter $\lamg$ can be determined from the difference between these two distributions (see the footnote for more details).}
\label{fig:2}
\end{center}
\end{figure}

We show examples of the $\omega$ distribution generated by the Monte Carlo (MC) simulation in Fig.~\ref{fig:2}.

Now we explain how to extract the value of $\lamg$ as well as its statistical error from a given $\omega$ distribution. We will we present our sensitivity study result in section \ref{sec:6}. Since the use of the $\omega$ variable reduces our fit to a one-dimensional one, $\lamg$ is obtained simply as a solution to the following equation: 
\be
\frac{\partial \ln {\mathcal{L}}}{\partial \lamg}=N\langle \frac{\omega}{1+\lamg \omega}\rangle =0.
\label{eq:19}
\ee


Of course, one could solve the equation~\eqref{eq:19} by successive searches. However, we can provide explicit expressions for $\lamg$. One sees easily that the normalised distribution in $\omega$, $W'(\omega)$, can be written as :
\bea
W'(\omega)=\phi(\omega)(1+\lamg\omega)
\eea 
where $\phi(\omega)$ is an \underline{even} function of $\omega$, since as can be seen from Eq.~\eqref{eq:fandg} $f(s_{13},s_{23},\cos\theta)$ is an even function of $\cos\theta$ while $\omega(s_{13},s_{23},\cos\theta)$ is an odd one.

Then, one can easily demonstrate by integration over the interval $-1<\omega<1$ that $\lamg$ can be expressed as ratios of odd over even momenta:
\bea
\lamg=\frac {\langle\omega^{2n-1}\rangle} {\langle\omega^{2n}\rangle} ~ ~ ~ (n\ge1)
\label{lamg}
\eea
Therefore, the expression obtained by DDLR for small $\lamg$ seems exact.

Similarly to Eq.~(\ref{eq:19}), one can also obtain the statistical error to the given value of $\lamg$ as: 
\be
\sigma^2_{\lamg} =\frac{1}{N\langle \left(\frac{\omega}{1+\lamg \omega}\right)^2\rangle}. 
\label{eq:21}
\ee
Thus, once the $\omega$ distribution is obtained experimentally, the Eqs. (\ref{eq:19}) or (\ref{lamg}) and (\ref{eq:21}) immediately provide the values of $\lamg$ and $\sigma_{\lamg}$. 
\footnote{In the real data, one must consider the systematic errors coming from detector effect etc and perform a $\chi^2$ fit instead of using these simple formulae. There is one subtlety for that case.  For each event, the photon should have the polarization either left- or right-handed. Thus, in MC, we produce the $\omega$ distribution with purely left- and right-handed PDF. Then, the total $\omega$ distribution of the experimental data is expected to be a linear combination of these two distribution with a ratio of $\epsilon$: 
\[
N^{\rm Exp.}(\omega)=\epsilon N_R^{\rm MC}(\omega)+(1-\epsilon)N_L^{\rm MC}(\omega)
\]
with $\epsilon\equiv\frac{1+\lamg}{2}$. The $N$ is the number of event in the experimental measurement. 
We show an example of the $\omega$ distribution of $\lamg=-1$ (red) and  $\lamg=+1$ (blue) in Fig.~\ref{fig:2}. 
As seen in this equation, the $\lamg$ can be determined from the difference between these two distributions.}

\section{Future prospects for the polarization measurement}
\label{sec:6}

In this section, we discuss the sensitivity of the future experiments, namely the SuperB factories and the LHCb to $\lambda_\gamma$, using the $B\to K_1(1270)\gamma\to (K\pi\pi)\gamma$.  We also discuss the advantages and disadvantages of our method compared to the other methods of the polarization measurement using the other processes, such as $B\to K^*e^+e^-$, $B_d\to K^*\gamma$ and $B_s\to \phi\gamma$.

\subsection{The sensitivity study of the polarization measurement with $B\to K_1(1270)\gamma$ in the DDLR method}

In this section, we perform a Monte Carlo simulation in order to estimate the sensitivity of the future experiments to the polarization parameter $\lambda_\gamma$ using the DDLR method. 
Following the procedure described in Section 5, 
we first generate the events ($10^3$ and $10^4$ events as examples) for a given value of $\lambda_\gamma$ and then estimate the expected statistical error, $\sigma_{\lambda_\gamma}$. Here, we use the ``ideal'' Monte Carlo simulation, i.e. detector and background effects are not taken into account. In order to generate the events as well as to compute the $\omega$ distribution, we use the input hadronic parameters as given in the Section~4, taking into account the form factor momentum transfer dependence (discussed in Appendix~B). These parameters include the experimentally measured isobar widths, the $^3P_0$ model parameters (the meson wave function radii, the quark-pair-creation constant, damping factor) and the phenomenological $K_1$ mixing angle.
In the Table~\ref{tab:lambda}, we present our result in the case of the SM, i.e. $\lambda_\gamma=1$. 
One can see from the table, that for $10^4$ events the error on $\lambda_\gamma$ is smaller than 0.1. We found that the errors do not change much for different values of $\lamg$.
We found that the $\omega$ distributions for the $K^{+}\pi^+\pi^-$ and $K^{0}\pi^+\pi^-$, and  $K^{0}\pi^{+}\pi^0$ and  $K^{+}\pi^{-}\pi^0$ are the same. Then, it should be pointed out an advantage of using the $\omega$-variable: all the channels corresponding to the same PDF can be merged altogether. That means that one can compute $\omega$-variable for each event and build a single histogram, which can increase the statistical significance.

In the above, we use the full decay distribution, not only on the information of the angular part but also the information of the invariant mass of the hadronic system. In the original DDLR paper~\cite{DDLR}, it was pointed out, using an average decay distribution in place of a full decay distribution for each set of invariant masses results in a decrease of the sensitivity. In order to test this,  we also produce the $\omega$ distribution including only the $\cos\theta$ dependence, i.e. integrated over the Dalitz plot, and compute  $\sigma_{\lambda_\gamma}$. We found that the inclusion of the full Dalitz information can indeed improve the sensitivity by typically a factor of two comparing to the angular fit.


Up to now, we have not considered the systematic errors coming from the hadronic parameters. 
We must reiterate that our hadronic model applied in the above analysis is approximate; it depends on basic assumptions like the non relativistic approximations inherent to the quark models. It depends also on parameters, some of them being internal to the full quark model, like the meson radii, and one being purely phenomenological, the mixing angle $\theta_{K_1}$ (we must note that there exists a correlation between the mixing angle, extracted from the data, and the chosen set of meson radii).
It depends also on the set of experimental data which we claim to describe by such models as discussed in Section 4. 
It is then a difficult question to evaluate the uncertainties of our results ; we do not claim to discuss this point precisely in this paper but we intend to do this in another publication. 


Finally, we would like to give a rough estimate for the event numbers expected by the future experiments, namely the SuperB factories and the LHCb. 
Taking the exclusive branching fraction $Br(B^+\to K_1^+(1270)\gamma)=4.3\times10^{-5}$ and assuming that the decays $K_1\to K\pi\pi$ are  by $K^*\pi$ (16$\%$) and $K\rho$ (42$\%$) channels, we obtain the observable branching fraction of $Br(B^+\to(K^+\pi^-\pi^+)_{K_1(1270)}\gamma)=4.3\times10^{-5}\times(0.16*4/9+0.42*1/6)\simeq0.6\times10^{-5}$ and $Br(B^+\to(K^0\pi^+\pi^0)_{K_1(1270)}\gamma)=4.3\times10^{-5}\times(2*0.16*2/9+0.42*1/3)\times1/3\simeq0.3\times10^{-5}$ (here the last factor $1/3$ comes from the fact that $K^0$ is observed as $\pi^+\pi^-$ from the $K_S$ decay). 
In order to get a more realistic estimation of the required number of signal events in the future experiments, we take the total efficiency of the reconstruction and selection to be of the order of 0.1$\%$ as in the case of $B\to K^*\gamma$ and $B_s\to\phi\gamma$ at the LHCb experiment~\cite{LHCb-Bphigamma} and of the order of 1$\%$ at B factories~\cite{Belle-BK1gamma}. Then, we obtain the yield of the nominal data taking to be of the order of $5\times10^3$ of $B^+\to(K^+\pi^-\pi^+)_{K_1(1270)}\gamma$ and $2.5\times10^3$ of $B^+\to(K^0\pi^+\pi^0)_{K_1(1270)}\gamma$ signal events for the accumulated luminosity of 2~fb$^{-1}$ at LHCb. The estimated annual yield at SuperB factories with 2~ab$^{-1}$ of integrated luminosity is of the order of $1\times10^3$ and $0.5\times10^3$ of signal events of $B^+\to(K^+\pi^-\pi^+)_{K_1(1270)}\gamma$ and $B^+\to(K^0\pi^+\pi^0)_{K_1(1270)}\gamma$,  respectively. 
Thus, the event sample, $10^3$ and $10^4$, studied in Table 1, roughly corresponds to the annual expected events of SuperB and LHCb, respectively. 
It should be noted that the decay modes including  a neutral particle is difficult to be studied in LHCb, i.e. LHCb may well study the first decay channel in the Table 1 while SuperB can study all of them reasonably well.

\begin{table}
\centering
\begin{tabular}{|c|c|c|}
\hline
$\sigma_{\lamg}$ (statistical error)& $N_{events}=10^3$ &$N_{events}=10^4$\\
\hline
\hline
$B^+\to (K^+\pi^-\pi^+)_{K_1(1270)}\gamma$ & $\pm$ 0.18 & $\pm$ 0.06 \\
\hline
$B^+\to (K^0\pi^+\pi^0)_{K_1(1270)}\gamma$ & $\pm$ 0.12 & $\pm$ 0.04 \\
\hline
\hline
$B^0\to (K^0\pi^+\pi^-)_{K_1(1270)}\gamma$ & $\pm$ 0.18 & $\pm$ 0.06 \\
\hline
$B^0\to (K^+\pi^-\pi^0)_{K_1(1270)}\gamma$ & $\pm$ 0.12 & $\pm$ 0.04 \\
\hline
\end{tabular}
\caption{Sensitivity study of the polarization measurement with  $B\to K_1(1270)\gamma$ in the DDLR method. Our estimates of  the statistical errors to $\lamg$ in the case of SM (i.e. $\lamg=+1$) is shown in this table. 
The event sample, $10^3$ and $10^4$, roughly corresponds to the annual expected events of SuperB and LHCb, respectively. 
The hadronic parameters used to obtain this result are given in section 4. The systematic error due to the uncertainties from these hadronic parameters is not included and has to be carefully studied.}
\label{tab:lambda}
\end{table}



\subsection{Comparison to the other methods}

In this subsection we compare the precision of the photon polarization measurement, using various direct and indirect methods.

\subsubsection{Comparison with the up-down asymmetry of GGPR}

One of the direct methods of the photon polarization determination methods, proposed by Gronau {\it et al.}~\cite{Gronau}, is to study the angular distribution in the $\overline{B}\to P_1P_2P_3\gamma$ decay and extract the polarization parameter $\lambda_\gamma$ from the angular correlations among the final hadronic decay products $P_1P_2P_3$. An observable called ``up-down'' asymmetry is introduced:
\begin{equation}
\mathcal{A}_{up-down} \equiv \frac{\int_0^1\,d\cos\theta\frac{d\Gamma}{d\cos\theta} - \int^0_{-1}\,d\cos\theta\frac{d\Gamma}{d\cos\theta}}{\int_{-1}^1\,d\cos\theta\frac{d\Gamma}{d\cos\theta}} = \frac{3}{4}\lambda_\gamma\frac{\int dsds_{13}ds_{23}Im[\vec{n}\cdot(\vec{\mathcal{J}}\times\vec{\mathcal{J}}^*)]}{\int dsds_{13}ds_{23}|\vec{\mathcal{J}}|^2}
\label{}
\end{equation}
representing the asymmetry between the measured number of signal events with the photons emitted above and below the $P_1P_2P_3$ decay plane in the $\overline{K}_1$ reference frame.
Having the theoretical prediction of $\mathcal{J}$, one can determine $\lambda_\gamma$.

Our conclusion, identical to the one for the angular fit, is that the statistical error on $\lambda_\gamma$ is about twice the one in our method.

\subsubsection{Comparison with $B \to K^*\ell^+\ell^-$}

From the analysis of the angular distributions of the four-body final state in the $B^0\to K^{*0}(\to K^-\pi^+)\ell^+\ell^-$ decay in the low $\ell^+\ell^-$ invariant mass region one can study various observables that involve different combinations of $K^*$ spin amplitudes~\cite{Kruger}.

Working in the transversity basis $\mathcal{M}_\perp = \frac{\mathcal{M}_R-\mathcal{M}_L}{\sqrt2}$ and $\mathcal{M}_\parallel = \frac{\mathcal{M}_R+\mathcal{M}_L}{\sqrt2}$, one the most promising observable, that has a small impact from the theoretical uncertainties, is the transverse asymmetry defined as
\be
\left.\mathcal{A}_T^{(2)}\equiv\frac{|\mathcal{M}_\perp|^2-|\mathcal{M}_\parallel|^2}{|\mathcal{M}_\perp|^2+|\mathcal{M}_\parallel|^2}=-\frac{\mathcal{M}_R\mathcal{M}_L^*+\mathcal{M}_R^*\mathcal{M}_L}{|\mathcal{M}_R|^2+|\mathcal{M}_L|^2}\right|_{\text{SM}}\approx -2\frac{\mathcal{M}_L}{\mathcal{M}_R}
\label{eq:AT12}
\ee
Note that we assume that $\mathcal{M}_{L/R}$ at the low $l^+l^-$ invariant mass regions can be identified with the decay amplitudes of $b\to s\gamma_{L/R}$  and are related to our polarization parameter as $\lambda_\gamma\simeq \frac{|\mathcal{M}_R|^2-|\mathcal{M}_L|^2}{|\mathcal{M}_R|^2+|\mathcal{M}_L|^2}$

The new analysis of the $B\to K^*e^+e^-$ decay mode by the LHCb collaboration~\cite{LHCb-BKstee} shows that one can expect an annual signal yield of 200 to 250 events for 2~fb$^{-1}$ 
in this energy region.  With this number, it is  found that 
the LHCb can reach a precision of $\sigma(A_T^{(2)})$ around 0.2 corresponding to the statistical error on $\sigma(\mathcal{M}_L/\mathcal{M}_R)$ to be of the order of 0.1~\cite{LHCb-BKstee}.

\begin{figure}[h]
\begin{center}
\epsfig{file=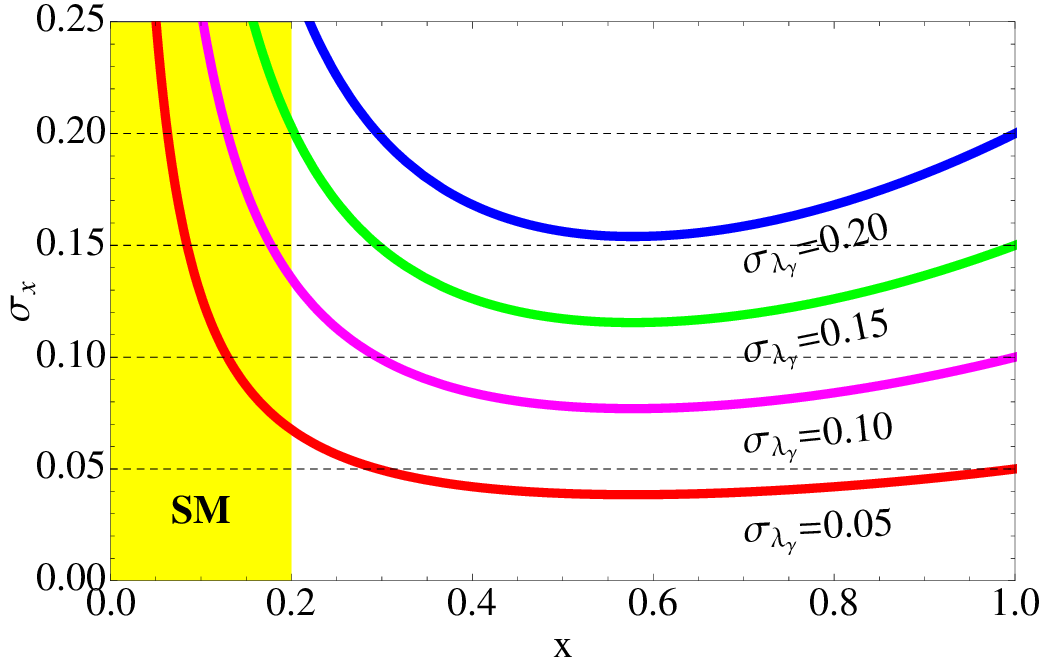,width=10.cm}
\caption{Comparison of the sensitivity of the two methods: the  one directly determining $x\equiv |\mathcal{M}_L/\mathcal{M}_R|$ and the other one determining $x^2$ such as our $\lamg$ (see Eq.~(\ref{eq:44})). One can see that when we assume the same errors for the both methods,  
a better significance can be obtained with the later method for $x\gsim 0.3$.}
\label{fig:sigmax}
\end{center}
\end{figure}

It should be noticed that this method allows the direct measurement of the ratio $x\equiv|\mathcal{M}_L/\mathcal{M}_R|$, while our polarization parameter $\lambda_\gamma$ is sensitive only to the amplitude ratio square, $x^2$. Therefore, the errors of these two methods are to be compared using the following equation:
\begin{equation}
\sigma_x=\frac{(1+x^2)^2}{4x}\sigma_{\lambda_\gamma}
\label{eq:44}
\end{equation}
which shows that the sensitivity depends on the value of $x$.  
We should immediately notice that for verifying the SM value, $x\simeq 0$, the method accessible to $x$ is much more advantageous than the one to $x^2$: our $\lambda_\gamma$ is in fact insensitive to the SM point (requiring an infinitesimal error). 
We plot Eq.~\ref{eq:44} in Fig.~\ref{fig:sigmax}. 
Let us look at the horizontal line of $\sigma_x=0.1$, expected error on $x$ with the $B\to K^*e^+e^-$ measurement. One can see that our method becomes more advantageous above $x\sim 0.3$, where the same sensitivity to $x$ can be achieved with a larger error to $\lambda_\gamma$ i.e. $\sigma_{\lambda_\gamma}>0.1$.

\subsubsection{Comparison with the methods invoking $CP$ asymmetries}

An indirect method to measure the photon polarization is to study the time-dependent $CP$ asymmetry in the neutral $B_q$ ($q=d,s$) mesons. 
For the generic radiative decay of the neutral $B_q$-meson into any hadronic self-conjugate state $M^{CP}$, $B_q(t)\to M^{CP}\gamma$, neglecting direct $CP$ violation and the small width difference between the two $B$-mesons, a $CP$-asymmetry is given by~\cite{Atwood}
\begin{equation}
\mathcal{A}_{CP}(t)=\xi\sin(2\psi)\sin(\phi_M-\phi_L-\phi_R)\sin(\Delta mt)
\label{eq:ACP}
\end{equation}
where $\xi(=\pm 1)$ is the $CP$ eigenvalue of $M^{CP}$, $\sin(2\psi)\equiv\frac{2|\mathcal{M}_L\mathcal{M}_R|}{|\mathcal{M}_L|^2+|\mathcal{M}_R|^2}$ parametrizes the relative amount of left- and right-polarized photons, $\phi_{L,R}=\sin^{-1}\left(\frac{Im\mathcal{M}_{L,R}}{|\mathcal{M}_{L,R}|}\right)$ are the relative $CP$-odd weak phases in the $b\to s\gamma$ process and $\phi_M$ is one in the  $B_q-\overline{B}_q$ mixing. 
These phases are $\phi_{L/R}=0$, $\phi_d=2\beta$, $\phi_s\simeq 0$ in SM. 
The smallness of the right-handed amplitude in SM, $\mathcal{M}_L/\mathcal{M}_R\simeq 2m_s/m_b$, predicts  $\mathcal{A}_{CP}(t)\simeq 0$. We should emphasise that  $\mathcal{A}_{CP}(t)$ measures the combination of $x\equiv |\mathcal{M}_L/\mathcal{M}_R|$ and the CP violating phases $\phi_{M,L,R}$ but not separately. Thus, the value of $x$ can be obtained from this measurement,  {\it only by having the value of the  CP violating phases in the $b\to s\gamma$ as well as the $B_q$ mixing}.

Current world average for the asymmetry in the $B_d\to K_S\pi^0\gamma$ process is $S_{CP}(B\to K_S\pi^0\gamma)=-0.15\pm0.20$~\cite{HFAG}, which is expected to be improved by the SuperB factory; error down to $2\%$~\cite{SuperB}.  
The LHCb experiment is going to measure $B_s\to \phi \gamma$ process. 
Based on the MC simulation for 2~fb$^{-1}$, it is claimed in~\cite{LHCb-Bphigamma} that LHCb will be able to measure $x$ with the accuracy of $\sigma_x\simeq0.1$. Therefore, similar to the case of $B\to K^*e^+e^-$, our method using $\lambda_\gamma$ can be more sensitive to $x$ above $x\sim 0.3$ (see Fig.~\ref{fig:sigmax}). Again, it should be emphasised that although an observation  $\mathcal{A}_{CP}(t)\neq 0$ in this method immediately indicates the existence of new physics, a quantitative determination of $x$ is not possible unless we fix the new physics model, namely the CP violating phases in $b\to s\gamma$ as well as $B_q$ mixing.

\section{Conclusions}
\label{sec:8}
We investigated the method to determine the photon polarization of the $b\to s\gamma$ process using the decay channel $B\to K_1\gamma\to K \pi \pi\gamma$, which was originally proposed by Gronau et al~\cite{Gronau}. In this paper, we propose a new variable, $\omega$, to determine the polarization parameter $\lamg$. This variable was firstly applied in the $\tau$ polarization measurement in the ALEPH experiment~\cite{DDLR}. The use of $\omega$ significantly simplifies the experimental analysis and as a result, it allows us to include not only the angular dependence of the polarization parameter, represented as the up-down asymmetry in~\cite{Gronau}, but also the three body Dalitz variable dependence to the fit. We found that when the data is analysed by using $\omega$, the statistical error in the polarization parameter $\lamg$ can be reduced by a factor of 2, comparing to the case of the up-down asymmetry.

In order to evaluate the systematic error, a sufficiently accurate modelling of the hadronic decays of $K_1\to K\pi\pi$ is required. Having the recent observation of the Belle collaboration~\cite{Belle-BK1gamma} implying the dominance of the $B\to K_1(1270)\gamma$ channel over $B\to K_1(1400)\gamma$, we investigated the hadronic decay of $K_1(1270) \to K\pi\pi$ in some detail. 
We first have derived  the basic hadronic parameters required in our analysis. These parameters can, in principle, be determined by the experimental measurements of the $K_1(1270)\to K \pi\pi$ decay. On the other hand, although the outstanding ACCMOR experiment had provided an extensive study of this decay, we found that the information one can extract from it is not accurate enough. We described some of the  problems encountered in our analysis, which include the strong phase between different intermediate resonance states and the controversial $K_1(1270)\to K\pi\pi$ through scalar mesons. Interestingly, the latter problem has been studied by the Belle collaboration recently~\cite{Belle-BpsiK1}, with a small result contradicting to the PDG number and in an agreement with our prediction by the $^3P_0$ model. Their results will provide a great help for our future study of the $K_1$ decay modes.

Being unable to obtain the hadronic parameters from the fundamental theory, we resorted to combine experimental data and phenomenological models. Practically, combining the experimental results of the partial wave analysis of the $K_1$ decays and the predictions of the $^3P_0$ quark-pair-creation model, we found that the $K_1$ mixing angle between $50^{\circ}$ to $60^{\circ}$ is well compatible with the experimental data.
Nevertheless, an evaluation of the theoretical uncertainties requires much more detailed discussions, which we will present in the forthcoming paper.

\section*{Appendix A: The $K_1\to K\pi\pi$ form factors}
In this appendix, we derive the quasi-two-body decay amplitude $\mathcal{M}_{(P_iP_j)P_k}^V$ given in Section~\ref{sec:4}. For the computation of this amplitude, we take into account the vector meson resonance width effect assuming the Breit-Wigner form, thus
\be
\mathcal{M}_{(P_iP_j)P_k}^V\equiv
 \mathcal{M}(K_1\to VP_k) \mathcal{M}(V \to{P_i P_j}){\rm BW}_V(s_{ij})
\ee
The decay amplitude of the axial-vector $K_1$ to a vector ($V$) and a pseudoscalar ($P_k$) meson can be expressed in the following Lorentz invariant form:
\begin{equation}
	\langle V(p_{V},\varepsilon^{(V)}) P_k(p_k))|\Delta H_A|K_1(p_{K_1},\varepsilon^{(_{K_1})})\rangle=\varepsilon_\mu^{({K_1})}T^{\mu\nu}\varepsilon_\nu^{(V)*}~,~~T^{\mu\nu}=f_Vg^{\mu\nu}+h_V p_{V}^{\mu}p_{K_1}^{\nu}
	\label{eq:AVP}
\end{equation}
where $f_V$ and $g_V$ are the form factors.
The amplitude of the subsequent decay $V$ to two pseudoscalar mesons $P_i$ and $P_j$ can be parametrized in terms of one vector-pseudoscalar coupling $g_{VP_iP_j}$:
\begin{equation}
	\langle P_i(p_i)P_j(p_j)|\Delta H_{V}|V(p_{V},\varepsilon^{(V)})\rangle=g_{VP_iP_j}\varepsilon_\mu^{(V)}(p_i-p_j)^\mu
\end{equation}
Using these form factors, we can obtain 
\be
\mathcal{M}_{(P_iP_j)P_k}^V =
(\vec{p}_i\cdot \vec{\epsilon}_{K_1}) a_{ij}^V+(\vec{p}_j\cdot \vec{\epsilon}_{K_1}) b_{ij}^V
\ee
where 
\be
\begin{split}
a_{ij}^V &= g_{VP_iP_j}{\rm BW}_V(s_{ij})[f_V+h_V\sqrt{s}(E_i-E_j)-\Delta_{ij}] \\
b_{ij}^V &= g_{VP_iP_j}{\rm BW}_V(s_{ij})[-f_V+h_V\sqrt{s}(E_i-E_j)-\Delta_{ij}]
\end{split}
\ee
with $\Delta_{ij}\equiv \frac{(m_i^2-m_j^2)}{M_{ij}^2}[f_V+h_V\sqrt{s}(E_i+E_j)]$. Note $E_i=(s-s_{jk}+m_i^2)/(2\sqrt{s})$. 

Finally, using these $a_{ij}^V$ and $b_{ij}^V$ functions, we obtain the $K_1\to P_1P_2P_3$ amplitude as: 
\be
\begin{split}
\mathcal{M}(K_1\to P_1P_2P_3) &=
c_{132} \mathcal{M}_{(P_1P_3)P_2}^V
+c_{231} \mathcal{M}_{(P_2P_3)P_1}^V
+c_{123} \mathcal{M}_{(P_1P_2)P_3}^V \\
&\equiv
(\vec{p_1}\cdot \vec{\epsilon}_{K_1})\see_1
-(\vec{p_2}\cdot\vec{\epsilon}_{K_1})\see_2
\end{split}
\ee
where
\be
\begin{split}
\see_1 &=
c_{132}(a_{13}^V-b_{13}^V)
-c_{231}b_{23}^V
+c_{123}a_{12}^V \\
\see_2 &=
c_{132}b_{13}^V
-c_{231}(a_{23}^V-b_{23}^V)
-c_{123}b_{12}^V
\end{split}
\ee

\section*{Appendix B: Damping factor}
In this appendix, we discuss the necessity of introducing the cutoff in the coupling vertices. When we compute the transition rates, we must take into account, in principle, the widths of the initial or final resonances; this is especially crucial for the transition rate of $K_1(1270) \to K \rho$, which is large, although it would be kinematically almost forbidden at the nominal values of the masses. A well-known and simple way to take widths into account is by integrating over the off-shell "masses", $p^2$, with the weight of the Breit-Wigner's. However, it is then found that the integrals will diverge for P or D waves, due to the $k^{2 l}$ factors, where $k$ is the decay momentum, if the coefficients are taken to remain constant. Of course, the reactions will in general provide natural limits of integration, for instance the spectrum studied by ACCMOR stops at $M_{K \pi \pi}=1.6~$GeV, but even that cut would give exceedingly large $P$ or $D$ wave contributions. In fact it seems that various indications hint to the necessity of a strong dynamical cutoff, or "damping factor", affecting for instance the Breit-Wigner shape (e.g. accurate studies of $\Delta(1236)$~\cite{Angela} or $K^*(890)$, see ref.~\cite{Aston}); the prototype of which are the Blatt-Weisskopf factors. The need for it is also shown by calculations of hadronic loops in the $^3P_0$ model~\cite{Silvestre-Brac}. One obtains a natural damping factor through the gaussian factors $e^{-\beta k^2}$:
\begin{equation}
	A^S\propto (3-\alpha k^2)e^{-\beta k^2}, ~ ~ ~A^D\propto \alpha k^2 e^{-\beta k^2}
	\label{}
\end{equation}
but $\beta\sim0.3$~GeV$^{-2}$ is found there to be quite too small. Following
ref.~\cite{Silvestre-Brac}, we introduce the empirical Gaussian cutoff $\exp(-\beta'k^2)$, renormalizing $\beta\to\beta+\beta'$ and we fix $\beta'\approx 3$~GeV$^{-2}$.

Now, to study the $K_1$ decays, we integrate the squared amplitudes over the $K_1$ and vector meson resonance states ($K^*/\rho$) invariant masses within the whole kinematic allowed region in ACCMOR, $[1.0;1.6]$~GeV; the integration on $K_1$ invariant mass does not depend too much on these limits for $S$ waves. It is not the case for the $D$ waves, but once the damping factor is introduced, the $D/S$ ratio becomes stable. The isobar ($K^*/\rho$) decay does not depend much on the damping factor.

\section*{\bf \normalsize Acknowledgments}
We would like to thank J.~Lefran\c cois and M.-H.~Schune, F.~Le~Diberder and L.~Duflot, D.~Bernard,  W.~Dunwoodie, for very useful discussions and information. We thank warmly Damir Becirevic for his encouragement and his very useful criticism. A.~Le~Yaouanc is indebted to his colleagues L.~Oliver and J.-C.~Raynal for constant discussions on the subject. Work supported in part by EU Contract No.   MRTN-CT-2006-035482, \lq\lq FLAVIAnet'' and
by the ANR contract ``LFV-CPV-LHC'' ANR-NT09-508531. 

\appendix


\begin{thebibliography}{99}

\bibitem{HFAG}
  The Heavy Flavor Averaging Group {\it et al.},
  arXiv:1010.1589 [hep-ex].

\bibitem{Misiak}
  M.~Misiak {\it et al.},
  Phys.\ Rev.\ Lett.\  {\bf 98} (2007) 022002
  [arXiv:hep-ph/0609232].

\bibitem{Gronau}
  M.~Gronau and D.~Pirjol,
  Phys.\ Rev.\  D {\bf 66} (2002) 054008
  [arXiv:hep-ph/0205065].
; 
  M.~Gronau, Y.~Grossman, D.~Pirjol and A.~Ryd,
  Phys.\ Rev.\ Lett.\  {\bf 88} (2002) 051802
  [arXiv:hep-ph/0107254].

\bibitem{Belle-BK1gamma}	
  H.~Yang {\it et al.}  [Belle Collaboration],
  Phys.\ Rev.\ Lett.\  {\bf 94} (2005) 111802
  [arXiv:hep-ex/0412039].

\bibitem{DDLR} M.~Davier, L.~Duflot, F.~Le~Diberder and A.~Roug\'e, Phys.\ Lett.\ B {\bf 306} (1993) 411

\bibitem{InamiLim}
  T.~Inami and C.~S.~Lim,
  Prog.\ Theor.\ Phys.\  {\bf 65} (1981) 297
  [Erratum-ibid.\  {\bf 65} (1981) 1772].


\bibitem{Khodjamirian}
  A.~Khodjamirian, R.~Ruckl, G.~Stoll and D.~Wyler,
  Phys.\ Lett.\  B {\bf 402} (1997) 167
  [arXiv:hep-ph/9702318].

\bibitem{Bosch}
  S.~W.~Bosch and G.~Buchalla,
  JHEP {\bf 0501} (2005) 035
  [arXiv:hep-ph/0408231].

\bibitem{Grinstein} 
  B.~Grinstein and D.~Pirjol,
  Phys.\ Rev.\  D {\bf 73} (2006) 014013
  [arXiv:hep-ph/0510104].

\bibitem{Matsumori} 
  M.~Matsumori and A.~I.~Sanda,
  Phys.\ Rev.\  D {\bf 73} (2006) 114022
  [arXiv:hep-ph/0512175].

\bibitem{Ball}
  P.~Ball and R.~Zwicky,
  Phys.\ Lett.\  B {\bf 642} (2006) 478
  [arXiv:hep-ph/0609037].

\bibitem{Everett}
  L.~L.~Everett, G.~L.~Kane, S.~Rigolin, L.~T.~Wang and T.~T.~Wang,
  JHEP {\bf 0201} (2002) 022
  [arXiv:hep-ph/0112126].

  \bibitem{Hatanaka}
  H.~Hatanaka and K.~C.~Yang,
  Phys.\ Rev.\  D {\bf 77} (2008) 094023
  [Erratum-ibid.\  D {\bf 78} (2008) 059902]
  [arXiv:0804.3198 [hep-ph]].

\bibitem{Lee}
  J.~P.~Lee,
  Phys.\ Rev.\  D {\bf 74} (2006) 074001
  [arXiv:hep-ph/0608087].



\bibitem{Chung} S.~U.~Chung, Spin formalisms, CERN 1971
	
\bibitem{Daum}
  C.~Daum {\it et al.}  [ACCMOR Collaboration],
  Nucl.\ Phys.\  B {\bf 187} (1981) 1.


\bibitem{Babar-BK1pi}
  B.~Aubert {\it et al.}  [BABAR Collaboration],
  Phys.\ Rev.\  D {\bf 81} (2010) 052009
  [arXiv:0909.2171 [hep-ex]].


\bibitem{Belle-BpsiK1}
  H.~Guler {\it et al.}  [The Belle Collaboration],
  [arXiv:1009.5256 [hep-ex].


\bibitem{LOPR}
  A.~Le Yaouanc, L.~Oliver, O.~Pene and J.~C.~Raynal,
  Phys.\ Rev.\  D {\bf 8} (1973) 2223.

\bibitem{Kokoski}
  R.~Kokoski and N.~Isgur,
  Phys.\ Rev.\  D {\bf 35} (1987) 907.

\bibitem{Godfrey}
  S.~Godfrey and N.~Isgur,
  Phys.\ Rev.\  D {\bf 32} (1985) 189.

\bibitem{Blundell1}
  H.~G.~Blundell and S.~Godfrey,
  Phys.\ Rev.\  D {\bf 53} (1996) 3700
  [arXiv:hep-ph/9508264].

\bibitem{Blundell2}
  H.~G.~Blundell, S.~Godfrey and B.~Phelps,
  Phys.\ Rev.\  D {\bf 53} (1996) 3712
  [arXiv:hep-ph/9510245].

\bibitem{TKL} Forthcoming article on the hadronic decays of the $K_1$'s, A.~Tayduganov, E.~Kou, A.~Le~Yaouanc.

\bibitem{Dunwoodie} W.~M.~Dunwoodie, private communication


\bibitem{LHCb-Bphigamma} S. Barsuk et al., ``Roadmap for the radiative decays of beauty hadrons at LHCb'', LHCb public note

\bibitem{Kruger}
  F.~Kruger and J.~Matias,
  Phys.\ Rev.\  D {\bf 71} (2005) 094009
  [arXiv:hep-ph/0502060].


\bibitem{LHCb-BKstee} J.~Lefrancois and M.-H.~Schune, CERN-LHCb-PUB-2009-008

\bibitem{Atwood}
  D.~Atwood, M.~Gronau and A.~Soni,
  Phys.\ Rev.\ Lett.\  {\bf 79} (1997) 185
  [arXiv:hep-ph/9704272].

\bibitem{SuperB}
``SuperB: A High-Luminosity Asymmetric $e^+e^-$ Super Flavour Factory Conceptual Design Report'', INFN/AE-07/2, SLAC-R-B56, LAL 07-15, March 2007.

\bibitem{Angela} A.~Barbaro-Galtieri, Proc. Erice Summer School (1971) p.581

\bibitem{Aston}
  D.~Aston {\it et al.},
  Nucl.\ Phys.\  B {\bf 296} (1988) 493.

\bibitem{Silvestre-Brac}
  B.~Silvestre- Brac and C.~Gignoux,
  Phys.\ Rev.\  D {\bf 43} (1991) 3699.



\end{thebibliography}
\end{document}